# EUSO IN SEPTEMBER 2004

Ph. Gorodetzky for the EUSO collaboration
PCC-College de France, IN2P3/CNRS, 11 place Marcelin Berthelot, 75005 Paris, France

**Abstract:** After a short comparison of cosmic ray observation from ground versus space, EUSO detector and its capabilities are described. The political situation is sketched.

**Introduction:** Extreme Energy Cosmic Rays (EECRs, above $10^{19}$ eV) are observed from earth by two methods. The first one, the ground array, like AGASA, is composed of charged particle detectors scattered on a surface: each is a few m$^2$, and are about a km away from each other. At these energies, the shower is long enough that for vertical showers electrons strike the ground. However for showers more inclined than 45°, only muons are observed. The energy of the EHCR is reconstructed by a specific algorithm taking into account the pattern of detectors hit in an event (how many outside a circle of radius the distance interdetectors versus how many inside). The other method, fluorescence observation as in the HIRES experiment uses optical (300 to 400 nm) telescopes detecting the $N_2$ fluorescence due to the ionization by the shower electrons. AUGER is building an experiment using both techniques, on a large scale (5000 km$^2$ and 4 full (180°) telescopes) to detect EECR either with the ground array alone (day and night) or in coincidence (only the night naturally).

Pros and cons of the methods: ground array works 24 hour a day. It uses a pattern reconstruction based on the structure of the shower, that is its development in $\pi^0$, $\pi^+$ and $\pi^-$. Up to now, classical hadron physics have been assumed. However, if one realizes that the energy density obtained in the center of mass is above 3 GeV/fm$^3$ for proton energies above $10^{18}$ eV, then quark deconfinement can happen through a phase change. In the phase diagram where temperature is plotted versus density. At maximum temperature, the fireball thermalizes and the reaction is much slower than at lower energies: transverse energy increases, hence the width of the shower. In a ground array, more particles In the pattern algorithm, more particles are going to be observed outside the circle, simulating a higher than true energy. This algorithm should be changed, but nobody knows yet how. If, on the other hand, one is situated at maximum density (zero temperature), then, according to Bjorken [1], chiral symmetry can beak, inducing at very high rapidities (the center of the pattern seen by the array) a ratio of $\pi^0$ to charged pions very different from 1/3 as seen in some CENTAURO events [2]. Certain rapidity intervals can be populated only by $\pi^0$s, observed as electrons, others only by charged pions observed as muons. Again here, the pattern can be interpreted in a way such as inducing a wrong energy of the primary.

The fluorescence method is immune to the eventual change of hadron physics at high energies. It looks in a calorimetric way to the electrons. This is why this method seems more robust than the other one. However, the light has to propagate through atmosphere, and many impurities can attenuate it.

AGASA has events, around $10^{20}$eV, which do not quite follow the supposed GZK cutoff (above the cutoff), while HIRES follows the cutoff indicates that HIRES finds the energy of the interesting events lower than AGASA does. It can be resolved [3] by assuming a mistake in energy calibration having a value of 20%. This could indicate that quark deconfinement takes place. One of AUGER aims is to solve that problem by looking at both energies on the same events. It is interesting to see that their preliminary data [5] on hybrid events show an energy calibration of 30%, with, again the ground array energy higher than the fluorescence one.

What is really needed is accurate measurements. The EUSO scheme: looking at fluorescence from above (the ISS at 400 km altitude, with a telescope having a 60° opening angle sees on the earth a circle of 500 km diameter (AUGER is 50) hence a total target mass of $2\times10^{12}$ tons (air), 2000 times bigger than the $km^3$ at south pole. Fluorescence is more accurate, and the light absorbtion by air impurities is smaller from above than from under the shower.
It is also worth mentioning that most of the space experiments yield more accurate results than their ground counterpart: Hubble telescope, ESA's Integral, AMS…

EUSO description: EUSO has finished with flying honors its phase A. However, for reasons explained later, EUSO is not yet accepted in phase B.
The US 2 m Fresnel lenses are made of ultra pure PMMA with diffraction grooves to reduce the size of the Point Spread Function (PSF). Calculations have been made for the use of CYTOP, the cladding material of plastic optical fibers, which is much less dispersive than PMMA.
The Japanese photomultipliers (Hamamatsu R8900-M36) have 36 pixels (6x6) with no dead space between them. A huge effort is made in Riken to increase the photocathode efficiency by using a GaAsP photocathode whose quantum efficiency is 0.4 (instead of 0.2 for Bi-alcali), but requires a larger wavelenght. By using a shifter and a dichroic mirror, the full efficiency becomes ~ 0.32, however, the actual lifetime of such a photocathode is yet only 1/50 of the bi-alcali one.
The support for the photomultipliers and their electronics is designed by France; The PMT repartition is on an X-Y scheme. The shape of the focal surface is spherical. Some 5000 PMTs make about 200000 pixels (each about a square km on the ground).
Electronics are digital (Italy) and Analogic (France). Prototypes have been built and ASICs are being designed.
Switzerland (Neuchatel) is in charge of the 2 wavelenghts LIDAR which will give the altitude of the Cerenkov reflection (ground or cloud, and then which height).
Finally, ESA (through its Alenia factory) has designed EUSO envelope.

**Acceptance:** Simulations have been performed, taking into account the parametrization of the shower development (Corsika), the photons production from the fluorescence (Kakimoto [5]), and Cerenkov effect, atmosphere transport (Lowtran7 which takes into account Rayleigh and Mie diffusions and ozone effects), optics (transfer and aberrations), photo-detectors (filters and efficiencies) and finally the trigger where the threshold and persistence can be set at will. Through these simulations, the acceptance is such that for a $10^{20}$ eV shower at 60°, the background noise is about 0.5 photo-electrons per µs, the shower max is about 10, and the Cerenkov pulse (reflection on water) is also about 10. The shower signal lasts for about 100 µs (50 at mid height). These numbers show a tremendous requirement on the PMTs and their base have to be especially studied to cope with the high rate and the low allowable power.
The acceptance varies with energy, according to the trigger efficiency. With the actual trigger designed for the low power, where the progression of the shower cannot be followed from pixel to pixel, the efficiency is zero under $2\times10^{19}$ eV and saturates at 0.9 at $1\times10^{20}$. When one multiplies this efficiency by the rate of EECR impinging on earth (in $E^{-2.7}$), EUSO counting rate will be zero from $10^{19}$ eV, rise to a flat maximum between $3\times10^{19}$ eV and $5\times10^{19}$ eV, and drop back to zero (with the $E^{-2.7}$ law so that it is essentially zero at $5\times10^{20}$ eV. See later in this document how this could change with a more powerful trigger.

**Angular resolution:** the reconstructed showers show an angular resolution that goes down from some 11° for showers at 5° (nearly vertical), to 4° at 35° and less than 1° for showers around 80° (nearly horizontal). All showers with an angle > 60° have an angular resolution better than 2°. This is important for neutrino detection, see later.

**Clouds:** Again, simulations shave been made in the case of standard clouds with the conditions of a 5% albedo (which gives the intensity of the Cerenkov pulse bouncing back to EUSO), a standard trigger and a standard acceptance (608000 km2 sr). For E = $5\times10^{19}$ eV, fluorescence alone will permit to reconstruct 5% of the showers (taken with 100% being the full sin2θ), Cerenkov alone: 3% and Fluorescence plus Cerenkov: 27%. These numbers become respectively 5%, 13% and 53% for E = $1\times10^{20}$ eV and 5%, 27% and 53% for E = $5\times10^{20}$ eV. Cerenkov and Cerenkov plus fluorescence could increase if a higher value of the albedo was taken.
In case of no clouds, the same values at E = $1\times10^{20}$ eV are 7%, 1% and a very attractive 86%.

**Duty cycle:** The duty cycle is here the time fraction usable for measurements. It depends only on the photon background. This background has 3 sources: a) the stars reflect on earth with an albedo value similar to the value taken for Cerenkov reflection, b) the moon and c), the sun. a) alone during night without moon and without clouds has been measured in a balloon experiment above the Mediterranean sea (BABY [6]) to have a value of 300 photons/m$^2$/ns/sr. The moon effect b) is such that for 13% of the time, moon is under the horizon and if the maximum value acceptable is 1/3 of the above 300 photons/m$^2$/ns/sr, then the duty cycle is slightly better than 25%. Now, when the sun c) is "on", no measurement is possible. This leads to a remark about storms and lightning: when lightning is studied, daytime light is the background. So EUSO has to think of a way to protect itself off when reaching a storm.
EUSO has a warranty to stay on the ISS for 3 years. However, as seen later, if the Japanese HTV is used in replacement of an unavailable NASA shuttle, there is no return on earth.
Within 3 years, EUSO should gather some 500 events at E = $4\times10^{19}$ eV which is under the GZK cutoff. Then, 2 possibilities: there is no GZK and EUSO should see 300 events at E = $1\times10^{20}$ eV and 10 at E = $1\times10^{21}$ eV. If there is a full GZK cutoff, EUSO should see 50 events at E = $1\times10^{20}$ and about 1 at $4\times10^{20}$ eV.
Last point on that chapter: the comparison with AUGER. We suppose that EUSO starts operating at the beginning of 2010 and AUGER is only AUGER south. Then, at the beginning of 2013, EUSO has 10 times the number of ground array AUGER events. It will have 200 times the number of hybrid AUGER events.

**Energy resolution, errors:** EUSO "end to end" simulation evaluates the contribution of different factors involved in the shower energy determination;
- mass of projectile (from p to iron): 13%
- Fluorescence yield 12%
- Density of air 4%
- Temperature of air 8%
- Efficiencies of photomultipliers 12%
- Light transmission of lenses 7%
- Light transmission of atmosphere 18%
- Angle of shower 2%

Then the energy resolution $\sigma_E$ for a proton at nadir with E = $1\times10^{20}$ ev is 10%.

Neutrinos, horizontal showers: If the shower length depends on the encountered air density and the fluorescence production is constant with altitude from zero to 20 km (this is due to the fact that $O_2$ in air acts as a quencher, and, the higher the altitude, the less $N_2$ there is, so less fluorescence, but at the same time the less $O_2$, so fluorescence inhibit gets reduced), then one finds out that the shower length (time width) depends only on altitude, as shown in fig.1 where a $10^{20}$ eV proton hits the atmosphere at coordinates X = 200 km, Y = 35 km (35 km "north" of the center of EUSO field in a certain coordinate system. Four altitudes have been taken: 15, 20, 25 and 30 km. A further illustration of the universality of this curve is that two extra points (E = $5\times10^{19}$ eV and $1\times10^{21}$ eV) fall right on the curve.

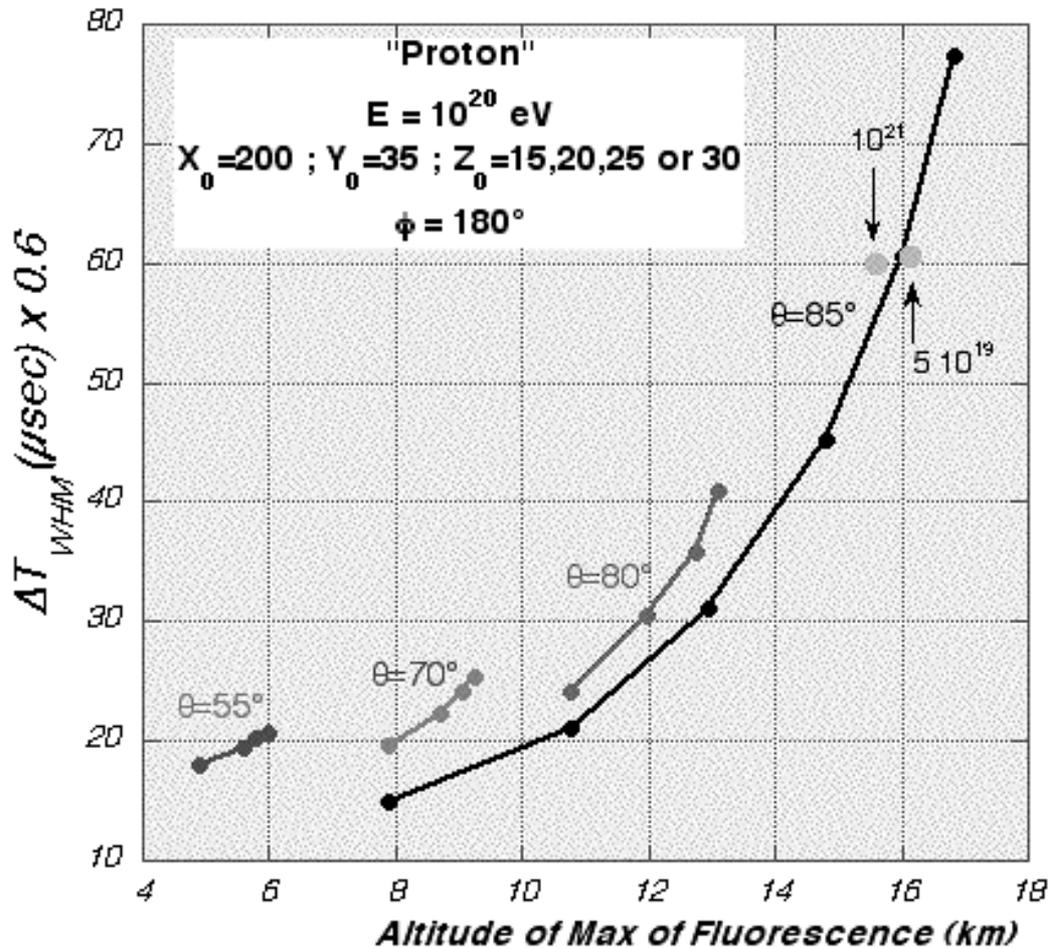

Fig. 1 Shower length versus altitude of shower max for different shower inclinations.

This can be applied to neutrino detection: a hadron horizontal shower can be seen only at around 20 km altitude (where the cone of vision of EUSO intersects the upper atmosphere), because hadrons interact immediately as soon as they encounter matter.
On the other hand, neutrinos have a very weak interaction cross-section. Hence, horizontal neutrinos will be detected with most probability at low altitudes (2 km) where density is high and even if they had to cross a few hundred km of air outside EUSO cone at that altitude, the probability to interact is constant.
If the only to sign unambiguously neutrinos is to see low lying horizontal showers, then AUGER cannot use its ground array. This means that in that field, EUSO is 200 times more efficient.

**EUSO political situation in September 2004.** Until June 2004, EUSO was in phase A. The ESA exam to terminate phase A was passed with flying honors. However, lack of money and the grounding of NASA shuttles pushed ESA Science Directorate to prevent EUSO to enter phase B. Microgravity and Manned Spaceflights (MSM) division at ESA seems to be ready to take the relay of the Science division in terms of finances. However, the local positions of

ASI in Italy, CNES in France is yet unknown. NASA has promised funding for the US collaboration (lenses) phase B, and has even begun to give money. However, the rest is linked to ESA agreement. In Japan, the Riken laboratory has agreed to pay for phase B (R&D on increasing the efficiency of photomultipliers) and, the miracle, JAXA has proposed to use their HTV launcher to replace NASA's shuttle. Drawings have been made, showing that EUSO can fit with its actual dimensions in the HTV. Then EUSO would dock on the ISS not anymore on the ESA platform Columbus, but on JAXA station. This would be a big advantage for the allowable power goes from actually some 1000 W to about 3000 W. Also, cooling would be provided by the docking station (EUSO does not need to cool itself anymore, which is simpler and a gain of space and weight). The increased power would be extremely useful to have better bases, holding very high rates for the photomultipliers. It could also allow to have embarked intelligence (a computer) allowing for a trigger which would follow in time and space the development of the shower from pixel to pixel. H. Crawford in Berkeley has proposed such a trigger, analogous to the Tpc trigger of the STAR experiment at RHIC. This trigger could lower the threshold in energy from $5 \times 10^{19}$ eV to $5 \times 10^{18}$ eV, allowing a much better overlap with AUGER. Finally, the HTV cannot go back to earth. So EUSO would stay on the ISS. The statistics of events gathered could increase.

**Conclusion:** EUSO is a mature project with high promises. It does not require new scientific developments, just assembling existing bricks. Naturally, if lenses can be made better, photomultipliers more efficient, trigger more intelligent, that should be done.

Looking at the fluorescence from space is a clean way to detect and measure showers. It is independent of any change in the physics of hadron collisions above a certain energy. Light transmission from above is excellent. It is certainly the tool of the future. Maybe now is too soon?